\documentstyle[12pt,amsfonts]{article}
\def\be{\begin{equation}}
\def\ee{\end{equation}}
\def\ba{\begin{array}}
\def\ea{\end{array}}
\def\d4{{\rm d}^4}
\begin{document}
\vskip -1.0cm
\title{\bf Gauge Gravity and Space-Time}
\author{ {Ning WU}\thanks{email address: wuning@ihep.ac.cn}
\\
\\
{\small Institute of High Energy Physics, P.O.Box 918-1, Beijing
100049, P.R.China} } \maketitle
 
\begin{abstract}

When we discuss problems on gravity, we can not avoid some
fundamental physical problems, such as space-time, inertia, and inertial
reference frame. The goal of this paper is to discuss the logic system of
gravity theory and the problems of space-time, inertia, and inertial
reference frame. The goal of this paper is to set up the theory on
space-time in gauge theory of gravity. Based on this  theory, it is
possible for human kind to manipulate physical space-time on earth, and produce
a machine which can physically prolong human's lifetime.
 
\end{abstract}
 
\newpage
\Roman{section}
\section{Symmetries in Physics}
It is found that basic physical laws of nature always show some
kinds of symmetry. For example, the classical Newtonian mechanics
has Galileo symmetry, Maxwell equations have Lorentz symmetry, and
electrodynamics has $U(1)$ gauge symmetry. On the other
hand, basic physical symmetries take an important role in the
construction of a physical theory. Sometimes, it even acts as an
important criteria for us to judge the rationality of a basic theory.
In a meaning, the symmetry in physics implies the symmetry of the law of nature.
Because of such symmetries of nature, a correct physical theory
must preserve the same symmetry. In physics, the symmetries
in basic physical laws play an important role in the construction
of a basic physical theory. The study of symmetry is an important
research field and method in the study of theoretical physics, especially in
the study of modern physics.
\\
There are mainly two kinds of symmetries in physics, one is the space-time
symmetry, and another is the internal symmetry. Space-time symmetry
is the symmetry that relates to the space-time coordinate
transformations, and
the internal symmetry is the symmetry of the transformation in the
space of internal degrees of freedom of a physical field without
changing space-time coordinates. In quantum field theory, the internal
symmetries in most cases are gauge symmetries. The space-time symmetries
are usually regarded as kinematical symmetries, while the internal
symmetries in most cases are dynamical symmetries. Besides, there is a
special symmetry, the gravitational gauge symmetry,
which have the property of both the space-time
symmetry and the internal symmetry. It is the gravitational gauge
symmetry, which study the local space-time translation symmetry
from the dynamical point of view\cite{01,02,03,04,05,06,07}.
\\
Symmetry means the invariance of basic physical laws under
mathematical transformations. From mathematical point of view,
the set of
these transformations forms a group. From the study of symmetries in
physics, we can see the beautiful mathematical structure of a physical
theory. Besides, we can study the conservation laws of nature.
It is known that, according to Noether's theorem, there is a conservation
law of the physical system which corresponds to each generator of the
symmetry group. For example, the energy-momentum conservation law corresponds
to the space-time translation symmetry, the angular momentum conservation
law corresponds to the rotational symmetry of space,
and the charge conservation law corresponds
to the electromagnetic $U(1)$ gauge symmetry. Noether's theorem reveals
the essential and positive connection between symmetry and conservation
law\cite{08}.
\\

\section{The Idea of Gauge}
\setcounter{equation}{0}
In modern physics theory, the most important role of symmetry
is not its relation to the conservation law, but that it provides a common
rule for the construction of a theory on interactions. It is known that,
soon after the foundation of quantum mechanics, the quantum mechanics
that describing electromagnetic interactions is found to be a gauge
invariant theory. After about 40 years, the unified electro-weak theory,
which is a $SU(2) \times U(1)$ gauge theory, is proposed. 10 years later,
the quantum chromodynamics  is founded. The quantum chromodynamics is
a $SU(3)$ gauge theory which describing strong interactions between
quark and gluon. An important inspiration that drawn from these
great physical achievements is that these three kinds of fundamental
interactions have a common nature, and are related to the same
physical rule. It is important to understand this point, for it provides
us a possible way to unify different kinds of fundamental interactions
of nature.
\\
Up to now, four kinds of fundamental interactions of nature are known
by human beings.
They are electromagnetic interactions,
weak interactions, strong interactions and gravitational interactions.
The first three are gauge interactions. An important idea that is drawn
from the studying of the first three kinds of fundamental interactions is
that the principle of local gauge invariance plays a fundamental role
in a physical theory on interactions, and it determines the forms of interactions.
So, a natural reaction of physicists is that gravity  should also be a
kind of gauge interactions, and the forms of gravitational interactions
should also be determined by the principle of local gauge invariance.
Studying along this direction, various kinds of gauge theory of gravity
are proposed\cite{a01,a02,a03,06}.
Comparing to the traditional gauge theories, there are mainly
two complexities for a gauge theory of gravity, one is that gravitational
field is a spin 2 gauge field, another is the influence from the idea of the
geometrization of gravity. Quantum gauge theory of gravity is proposed
based on the gauge principle\cite{01,02,03,04,05,06,07},
and the traditional geometrization idea
is not used in this theory. It is a perturbatively renormalizable quantum
theory, and it can returns to general relativity for classical
phenomenon of gravity. The  quantum gauge theory of gravity is proposed
based on the conviction that four kinds of fundamental interactions
of nature should have common rule, and they can be unified based on
this common rule\cite{09,10,11,12}.
\\
The core idea of gauge theory is gauge principle, which sets up the
essential and inexorable relation between symmetry and interactions.
In other words, it helps us to set up a systematic method to study
interactions based on its symmetry.
In gauge theory, when we determine a gauge symmetry, we can systematically
set up the corresponding theory of interactions, and all details of
the whole theory are almost uniquely determined by the gauge symmetry.
Conversely, if there exists a kind of interactions, we can determine
the corresponding gauge symmetry by gauge principle. A kind of interactions
must have its source, and this source must be a conserved charge.
According to Noether's theorem, this conserved charge must be
the generator of a symmetry group. This symmetry group is the
gauge group of the corresponding interactions. Based on this
gauge group, we can set up the gauge theory of the interactions.
But we know that, symmetry and interactions are two completely
different concepts of physics. Symmetry is a word that describes
static objects, while interactions are dynamic processes; symmetry
is a word that describes the global feature of a physical system,
while interactions mean changing in the local part of the system;
symmetry is the language of geometry, while interaction is the
language of physics. Though they are essentially different
concepts, they are unified based on gauge principle. The heart and
soul of quantum gauge theory of gravity is gauge principle.
\\

\section{Active Transformation and Passive Transformation}
\setcounter{equation}{0}

It is known that, in order to set up a gauge theory of fundamental
interactions, the first thing we need to do is to determine
the corresponding gauge transformation and
gauge group. For gauge theory of gravity,
the  gauge transformation is space-time
translation, for energy-momentum tensor is the generator of
space-time translation. But there are two ways to perform space-time
translation, one is active transformation, and another is
passive transformation. In fact, for other kinds of transformations,
such as Lorentz transformation and general coordinate
transformation, these two ways of transformation still
exist. For classical problems, active transformation
and passive transformation are mathematically equivalent.
Extending these mathematical equivalences to physics, we can
obtain two important ideas of physics, they are the relativity
principle and the equivalence principle. If what
we discuss is Lorentz transformation, then the active transformation
is the transformation that the physical object is boosted
while our observer keeps still, which corresponds to the
point of view of moving system; and the passive transformation
is the transformation that our observer is boosted while
the physical object keeps still, which corresponds to the
point of view of moving observer. If we make them equivalent
in physics, we will obtain the principle of special relativity,
for this equivalence in physics means that fixed reference
system and moving reference system are equivalent, which
means that all inertial reference system are equivalent, for
all of them are equivalent to the fixed reference system.
If what we discuss is general coordinate transformation,
the active transformation is the transformation that the
physical object is boosted arbitrarily while our observer
stands still, and our observer is an observer in an
inertial reference system; and the passive transformation
is the transformation that our observer is boosted arbitrarily
while the physical object is fixed. In the later case, our
observer is in a non-inertial reference system. If we make
them equivalent in physics, we will find that all observers,
no matter they are inertial or non-inertial, are equivalent.
In fact, this is a part of the principle of general covariance.
If our observer is in a local inertial reference system, this
equivalence means that an inertial reference system and a local inertial
reference system are equivalent, which is the idea of the
equivalence principle.
\\
However, active transformation and passive transformation have
different pictures in physics. Active transformation is the
transformation that is made directly on the physical system.
When we make local transformation, we will change structure
or states of the system. So it is a physical operation on
the system. In another words, it is a physical transformation.
Passive transformation
is a coordinate transformation, it is a mathematical transformation.
It does not cause any changes on the structure or states of the
system. In other words, the symmetry of active transformation
is an internal symmetry, while the symmetry of passive
transformation is a space-time symmetry. From theoretical point of
view, active transformation and passive transformation are
only two different ways that we mathematically describing
a transformation, but the physics that we study are the same.
In this meaning, they must be equivalent. However, though
they are equivalent, they may let to different physical
theories. Let's return back to the theories of gravity.
If we use the point of view of passive transformation in our study,
the local space-time translation is just the general coordinate
transformation. It is known that the best way to study
the  general coordinate transformation of a physical field
is to use the tensor theory. In order to keep the invariance
of a physical system in general coordinate transformations,
we need to introduce the concept of curved space-time. And after
the introduction of the concept of curved space-time, the theory
is complete, and there is no space for us to introduce
the concept of gravitational
field into the theory. A basic physical idea generated in this
study is that gravity is only an effect of the curved space-time\cite{13,14,15,16}.
So, in general relativity, the concept of curved space-time is
fundamental, it is an inevitable result of the theory. The concept
of gravitational field is a subsidiary concept of curved space-time.
But if we use active transformation in our study, local space-time
translation is just the gravitational gauge transformation. In this
case, space-time keeps unchanged, while physical field undergos
some transformations. We use the language of gauge transformation
to describe these changes. In order to keep the invariance of
our physical system under gauge transformations, what we need to do
is to introduce the corresponding gauge field into our theory.
This gauge field is just the gravitational gauge field. In this
case, we have no chance to introduce the concept of curved space-time
into our theory. So, in gauge theory of gravity, the concept
of gravitational gauge field is fundamental, and the concept
of curved space-time can not be introduced into the theory.
So, when we study the same physical problem, different
treatments in mathematics will cause great discrepancies on thoughts
of physics.
\\

\section{Absolute Space-Time and Physical Space-Time}
\setcounter{equation}{0}
In classical limit, quantum gauge theory of gravity can return to
general relativity. So, in a meaning, we can say that, for classical
problems of gravity, two theories are equivalent mathematically.
But two theories are not equivalent physically, for they have different
ideas on space-time. In general relativity, space-time is curved, and
gravity is only an effect of curved space-time. In general relativity,
there does not exist the concept of absolute space-time. In gauge theory
of gravity, gravitational field is a physical field which exists in space-time. So,
gravity and space-time are two concepts which are independent
each other. The independence of
gravity and space-time means that there should exist a space-time
which is not affected by gravity. This space-time is the absolute space-time,
which is always flat. In gauge theory of gravity, the absolute
space-time is a theoretical tool which is introduced transcendentally.
It is independent of matter, in other words, it will undergo no
changes when there are matters exist in it. It is a transcendental
basis of gauge theory of gravity.  But the space-time used by
us in the real world is the physical space-time,
because the space-time of our real world is affected by matter in it.
In other words, the so called time and space defined in physics are
affected by classical gravitational field. The time and space used by an
observer in an experiments are physical time and space, not the time
and space of absolute space-time. The physical time and space are
variable, and their changes are mainly caused by classical gravitational
field. Let's study how time is defined in physics. It is known that the time
itself is not directly perceptible, we feel the existence of time by
various effects of motion. In physics, the time is defined by effects
of motion. We have many methods to define the time in physics.
For example, the old-fashioned pendulum clock defines the time through
the period of  vibration of a pendulum, the traditional mechanical
clock defines the time through the period of vibration of a spring
oscillator, quartz clock defines the time through the period of vibration
of a quartz crystal, and atomic clock defines the time through the
period of oscillation of atomic spectrum. In all these methods, what
are used to define the time are effects of motion, not time itself.
Once classical gravitational field is introduced into space-time, these
effects of motion will be changed, and all of them will be changed
at the same manner. This result hold both in general relativity and
gauge theory of gravity. It is known that classical gravitational field
will put the influence to all motions of matter and interactions
at the same manner and the same magnitude, no
matter they are atom, molecular, quark, or gravitational field itself.
These influences are the so-called time dilation effect and
length contraction effect. So, when classical gravitational field
is introduced into space-time, the vibration period of a pendulum, of
a spring oscillator, of a quartz crystal, and of atomic spectrum will be
changed the same manner and the same  magnitude. Because the
changes of them all have exactly the same step, there do not
exist any relative changes when we compare these changes each other.
Therefor, local observer can not feel any changes of them. Local observer
makes a conclusion that, though the classical gravitational field of
surroundings is changed, the frequency of his clock undergoes no changes.
But a remote observer will find that the frequency of the clock used
by the local observer is changed. This is just what the gravitational
red shift experiment tells us. For the remote observer, the classical
gravitational field of his surroundings are very weak, so the time
given by his clock is approximately equal to the absolute time.
\\
Though the space-time of our real world is not absolute space-time,
the introduction of absolute space-time into quantum gauge theory
of gravity is necessary and inevitable. In quantum gauge theory of gravity,
in order to study quantum effects of gravitational field and the influence
of gravity to clock and  ruler, we need a space-time
frame which is not affected by gravitational field. On the other hand,
in order to study the law of the changes of time and space of physical
space-time, the introduction of absolute space-time is necessary.
Now, in order to grasp the concepts of absolute space-time and physical
space-time, let's do an ideal experiment. Suppose that there is an gravity
machine, which will be called Magic Gravity Box(MGB). We suppose that
MGB is a hi-tech machine that can change the magnitude of gravitational
potential inside MGB without any influence to the gravitational
potential outside MGB. We suppose that the gravitational potential
inside MGB is uniform, and the spacial derivative of gravitational
potential is small, so people in it will not feel very strong gravity.
Now, let's start our ideal experiment with our MGB. In the first step,
let's return back to the time of about 500,000 years ago. At that time,
a group of apes live near Beijing. They live in a huge hole in the Zhoukoudian
mountain. One day, they return back to the hole after hunting. They put
their stone knife and stone axe into their hole home, then they
roast their preys on fire and feast themselves in front of the hole.
After that, they are singing and dancing to celebrate their success
in hunting. Finally, they return their hole and sleep except one ape.
Only one ape was asked to go into the MGB to experience magic life in it.
After he go into the MGB, we adjust parameters of the MGB so that
the parameter $(1 - g C^0_0)$ inside it is about 500,000.
In the next day, Other apes awake after sunrise. They find that one
of their partner is missing. They agitate and search him everywhere.
Finally, they find him in a box with one meter long. They find that
the ape in the MGB is as small as a grain of dust, his heart stop
beating, and everything in the MGB keeps motionless. However, the ape
in the MGB does not feel that he undergos any changes. His heart keeps
beating as before, and his height is also the same as before. But he
is flabbergasted when he observes outside world through the window
of the MGB. His partners become as large as a huge mountain, and the
fine hair of his partners is huger than a big tree. He find that
the box is very huge, and the distance from the most east end to
the most west end is about 50 kilometers. When his heart beats every time,
there are six sunrise and sunset outside. He live in the MGB for
about one year. He feel quite loneliness, and he ask to let him out
to find his partners. Then, the parameter $(1 - g C^0_0)$
is adjusted to 1 to let him out. After he goes out, the first thing
he want to do is to find his partners in the hole, for he miss them
very much. But he find that everything was changed greatly. The hole
he ever lives in disappears, the small river that flows past the hole
also disappears, and the mountain is also changed greatly. None of
his partners is found. He falls into a depression, and walks hit or miss.
Suddenly, he found that there is a museum, and he walks into it. He
find the osseous remains of his partners become fossilized and is
displayed there. The stone knife and the stone axe, which are used by
him one year before, are also displayed there as rare treasures.
He is full of agony, for he has lost all partners, including his friends
and all his relatives. He dashed around like a bat out of hell.
He run into a huge city which is built of steel and cement. The
city is full of people and an Olympic game is held there. Some people
find that an ape is running in the street and report it to the police.
Finally, the ape is catched by policemen. Archeologists find that
the ape looks like Zhoukoudian ape. So, scientists are ask to
do DNA identification. DNA identification shows that he is just
the Zhoukoudian ape. Scientists can not understand how can he
pass through 500,000 years time and still alive now.
\\
The above description looks like a ugly science fiction. But here, MGB
is used to help us understand the great difference between the space-time
inside MGB and that out of MGB, which correspond to the physical space-time
and absolute space-time respectively. It should be stated that MGB
is not a pure fictitious machine. Of course, both classical Newton's theory
of gravity and Einstein's general relativity do not allow us to built
such a machine on earth. But, some quantum effects of gravity may help
us to built such a machine on earth, and quantum gauge theory of gravity
can provide us a theoretical guide on this study. Here, we will not go deep
into the theory and technology of MGB, which will be discussed in other papers.
Now, let's return back to the concept of space and time. Outside MGB, the
gravitational field is very weak, so the space-time used by the observer
outside MGB is approximately absolute space-time. However, inside MGB, the
gravitational potential is very strong, the clock and ruler used by the
observer inside MGB strongly affected by the gravitational potential inside
MGB, so the space-time inside MGB is a physical space-time. From the point of
view of an observer in absolute space-time, the time interval between ticks of a clock
and length of a ruler will be changed when gravitational potential is changed, but
the structure of space-time itself undergoes no change. From the point
of view of an observer in physical space-time, the time interval between ticks
of a clock and length of a ruler undergo no change when gravitational potential
is changed. But this invariance is only a relative invariance, for their
absolute magnitudes are changed.
\\
In general relativity, the physical gravitational field is not transcendentally
introduced, for gravitational field appears as an effect of curved space-time.
In general relativity, physical space-time is a transcendental concept. It is
not necessary to introduced the concept of absolute space-time, and there is no
existential space for the concept of absolute space-time. It is known that the
concept of absolute space-time is not consistent with the logic system of
general relativity, for the existence of the concept of the absolute space-time
need an implicit transcendental hypothesis that the existence of gravity is
independent of space-time, or gravity is not an internal structure of space-time.
From the point of view of quantum gauge theory of gravity, physical space-time
is a composite existence of absolute space-time and gravitational field. Therefore,
if physical space-time is treated as a transcendental concept in a theory,
neither absolute space-time nor physical gravitational field can be treated
as transcendental concepts. The logic system of quantum gauge theory of gravity is quite
different. In gauge theory of gravity, absolute space-time is a transcendentally
introduced. It is a logic starting point of the theory. Gravitational field
is a physical field existing in the absolute space-time. It is independent
of space-time. In other words, it is an independent physical field, not
an internal part of space-time. In gauge theory of gravity, the influence
of gravity to clock and ruler is explained by the point of view of interactions
between matter and gravitational field. What that gravity is treated as
an independent physical field in gauge theory of gravity requires that
the concept of space-time and the concept of gravity should be independent
each other. In other words, when we define space-time, or define clock
and ruler of space-time, there should exist no physical field in space-time.
In this case, the space-time is absolute space-time. So, general relativity
and quantum gauge theory of gravity have great differences on the concept
of space-time.
\\
 
\section{Physical Picture of Gravity and Geometric Picture of Gravity}
\setcounter{equation}{0}
Some people may ask such question like that space-time is flat or curved,
or the universe we live in is flat or curved? If it is flat, it can not
be curved, or if it is curved, it can not be flat. For example, you hold a
ball on your hand, no people will say that the surface of the ball is flat.
But if you hold a plate glass on your hand, no people will say that its
surface is curved. So, according to our ordinary experience of life,
it is indeed true that if something is flat,it can not
be curved, or if it is curved, it can not be flat. But for our universe,
its space-time structure is flat or curved? In order to answer this
question, we need to set up the representation theory of gravity. In fact,
similar questions exist in physics. Let's return back to the Galileo's
era of about 500 years ago. At that time, we faced a similar question.
Suppose that there is a table placed in a small cabin of a ship which
is sailing slowly in the sea. Our question is that the table is at rest
or is moving? If you sit in the boat, you will say that the table is at
rest; but if you stand in the seabeach, you will say that the table is moving.
Two answers are completely contradict. Which answer is correct? Galileo
study this problem, and two revolutionary ideas are generated from the study,
the relativity principle and the inertial principle. According to Galileo
principle of relativity, all motions are relative. If you want to describe
a motion, you need first to determine an observer, or select a reference
system. In order to answer the question that the table is at rest or is moving,
you need first to give a clear definition to the concepts of rest and motion,
or say that you need first select a reference system. A reference system
can be regarded as a representation of kinematics, and we will obtain
different results in different representations of kinematics. Next, let's
return back to the time of about 100 years ago. At that time, Heisenburg
found matrix mechanics. In matrix mechanics, operators of physical
quantities change with time, while quantum states are fixed. Soon after,
Schrodinger found wave mechanics. In wave mechanics, quantum states change
with time, while operators are fixed. It is found that both theories can
explain almost all quantum phenomena observed in experiments
at that time, but hypothesis
of two theories are completely contradict. Physicists want to know which
theory is correct? But final answer is exceeding all people's expectation,
that  is, both theories are correct. The key point to understand it is the introduction
of the concept of the representation of quantum theory. Now, we know clearly
that the wave mechanics founded by Schrodinger is the Schrodinger picture
of quantum theory, and the matrix mechanics founded by Heisenburg is the
Heisenburg picture of quantum theory. Two pictures are finally equivalent.
Two pictures study the same microscopic phenomenon by using different
mathematical tools from different point of view, but obtain the same final results.
But what is the exact reason that cause the divergence of two pictures?
Essentially speaking, the reason originates from the question of active
transformation or passive transformation, in other words, the question
of the transformation of system or the transformation of observer.
It can be called the relativity principle of transformation. We can make
such a correspondence: a wave function of quantum states can be regarded
as a basis of quantum states, which corresponds to the coordinate system
in classical mechanics, and an operator can be regarded as a quantum
object, which corresponds to the matter system in classical mechanics.
The developing of a quantum system can be describing in two ways. One way
is to adopt the idea of active transformation, that is the wave functions
are fixed while the operators are changing with time, which is just the
Heisenburg picture of quantum theory. Another way is to adopt the idea of
passive transformation, that is the operators are fixed while the wave
functions are changing with time, which is just the Schrodinger
picture of quantum theory.
\\
Now, let's return back to the topic of gravity theory. In fact, we face
similar questions. It is known that, in traditional gauge theory, the
conserved charge given by global gauge symmetry is just the source of gauge
interactions. We suppose that this rule holds for gravity theory.
We know that the source of gravitational field is energy-momentum, and
energy-momentum is the conserved charge of space-time translation
symmetry. Therefore, the symmetry of gravity theory is the space-time
translation symmetry. There are two mathematical ways to describe
space-time translation, which will generate two different representation
of gravity theory. One way is to use the manner of passive transformation.
In this way, physics system is fixed while space-time coordinates undergoe
translations. After localization, the localized space-time translations
are just the general coordinate transformations. In order to keep invariance
of gravity theory under general coordinate transformations, arbitrary
space-time metric and the concept of curved space-time are introduced
into the theory. It is found that, when the concept of curved space-time
is introduced into the theory, an independent gravitational field can not be
introduced into the theory, for the effect of curved space-time is just
gravity. Studying in this way, what we obtained is just general relativity.
In this theory, the concept of curved space-time is transcendental and
fundamental, and gravity is only a subsidiary effect. General relativity
has clear geometrical features, so we call it geometric picture of gravity.
Another way is to use the manner of active transformation. In this method,
space-time coordinates are kept unchanged while physical system undergoes
some transformations. This is the manner that we usually used to describe
symmetry transformations in gauge field theory. In gauge field theory,
what we study is gauge transformations, i.e., the gravitational gauge
transformations. In order to keep local gauge symmetry of the theory,
we need to introduce gravitational gauge field, which transmits gravitational
interactions. Because we perform no operations on the structure of space-time,
the structure of space-time are kept unchanged under gravitational gauge
transformations. Therefore, in quantum gauge theory of gravity, space-time
is always flat, or say that we have no reason to introduce the concept
of curved space-time into the theory. So, in quantum gauge theory of gravity,
the concept of absolute space-time is transcendental, whose existence is
independent of gravity, and gravity is a physical field which exists
in absolute space-time. In other words, space-time and gravity are
two independent concepts in quantum gauge theory of  gravity. The theory
constructed in this manner is physics picture of gravity. In geometrical picture
of gravity, space-time is physical space-time, so it is a curved space-time;
while in physics picture of gravity, space-time is absolute space-time,
so it is flat.
\\

\section{Clock and Ruler}
\setcounter{equation}{0}

Next, let's study the problem from a more fundamental point of
view. An obvious question is that, which mechanism
causes space-time curved? In other words, how to understand the
mechanism of space-time curving?
Or say that, from physical point of view, why
space-time in general relativity is curved, while it is flat in gauge theory
of gravity? Suppose that we are mathematicians. In this case, we just study
mathematical problems from mathematical point of view. We can make any
hypothesis on the structure of space-time without considering the structure
of space-time
of our real world. Suppose that there is a space-time. Before we make any further
hypothesis on the structure of space-time, it's structure is unknown, or say
that, it can be either flat or curved. Once the metric of the space-time is determined,
the structure of the space-time is determined at the same time. So, we conclude that
the structure of space-time is determined by its metric. But, what is the physical
nature of metric? From physical point of view, the key role of metric
is to define clock and ruler. Different definitions of clock and ruler will give out
different structures of space-time. The difference between different definitions of
clock and ruler in physics is that the exterior physical environments are different.
In physics, we have very strict restrictions on the
definitions of clock and ruler, and the purpose of
these strict restrictions is try to reduce the influence from exterior environments.
For example, the temperature of the instrument should be precisely set to a
given value, and the strength of electric field and magnetic field should be
small enough. It is known that both the time interval between ticks of a clock
and the length of a ruler are affected by temperature, strength of electric field and
magnetic field. They are also affected by gravitational field in space-time, but in
the definition of clock and ruler, we make no restrictions on the strength of
gravitational field, which causes different definitions of clock and ruler in
physics and different structures of space-time. It is known that the time interval
between ticks of a clock and the length of a ruler are affected by gravitational
potential, which are the famous time dilation and length contraction effects.
In order to avoid interference from gravitational potential, a natural requirement
is that, in the standard definition of clock and ruler, the gravitational potential
should be zero. When we adopt this definition, we will find that our universe
is flat, and our space-time is absolute space-time.
On the other hand, though gravity can change
the time interval between ticks of a clock and the length of a ruler, for a local
observer, all these changes are not observable, for in a local reference system,
all physical processes undergo the same changes which keeps all relative magnitudes
unchanged. For a local observer, all measurements are to determine the value
of relative magnitude, not of absolute magnitude. Though gravity can change
the time interval between ticks of a clock and the length of a ruler, a local
observer can not feel these changes. In other words, the local observer can not
find that his clock and ruler undergo any changes. For the sake of convenience,
he would like to use his clock and ruler to define time and space without
any modifications from gravity. The manner that to use the clock and
ruler in a local inertial reference system to define time and space  is used
by general relativity. If we adopt this manner to define time and space, we will
find that our universe is curved and our space-time is physical space-time.
The key mechanism that make our universe curved is that we use a clock
and a ruler which have different absolute magnitude in different point of
space-time. Therefore, that our universe is flat or curved is not determined
by existence, but by the clock and ruler selected by physicist. In this meaning,
the role of the equivalence principle can be regarded as to define clock
and ruler for a local system, or say that, it make the clock and ruler in
a local inertial reference system equivalent to the clock and ruler in an
inertial reference
system. The clock and ruler defined in this manner contains the effects of gravity.
Therefore, the effects of gravity are naturally put into space-time metric, and
gravity becomes a part of inner structure of space-time. So, the definition of
clock and ruler in general relativity is empirical, which is a natural selection
from the point of view of a local observer, while the definition of clock and ruler
in absolute space-time is transcendental, which is a natural selection from
the point of view of a theorist.
\\
General relativity and quantum gauge theory of gravity have quite different
theoretical structures and mathematical formulations. In fact, all these differences
originate from their different transcendental hypothesis. Some of these hypotheses
are not clearly mentioned in the standard formulation. We can only find them in the
implications of traditional formulations. These hypotheses contain the hypothesis of
the nature of gravity, the hypothesis on the nature of space-time, the hypothesis
on the nature of inertial, and the definitions of clock and ruler. The theoretical
foundation of general relativity is the equivalence principle. The nature of the
equivalence principle is to set up the equality of a local inertial reference system
and an inertial reference system, and the nature of this equality is to set up another
equality of the clock and ruler in a local inertial reference system and the clock
and ruler in an inertial reference system. The nature of the second equality is to
put the effects of gravity into space-time metric, and therefore to make gravity
to be a part of space-time metric. In this case, our natural results are that the
curved space-time is a basic and transcendental concept, and gravity is a subsidiary
concept, which is only an effect of curved space-time. Quantum gauge theory
of gravity inherits basic ideas and treatments of quantum field theory. In quantum
gauge theory of gravity, gravity is treated as a kind of physical interactions which exists in
space-time. What implicates in this manner of treatment is the hypotheses that
gravity and space-time are two transcendental concepts, and they are independent
of each other. Since gravity and space-time are independent of each other, or say
that gravity is an exterior object of space-time. In the definition of clock and ruler,
our natural requirement is that the local gravitational potential should be zero. Or
say that, the clock and ruler in a local inertial reference system can not be used to
define time and space. In this case, our space-time must be absolute space-time.
It can not be curved, for there does not exist any mechanism that can make space-time
curved. In fact, when we try to construct a physical theory, we can always supposed that our
space-time is transcendentally flat, for we can always take out all possible mechanisms
that can make space-time curved from the inner structure of space-time. These mechanisms
can be treated independently as an independent physical process.
\\

\section{Inertial Reference System}
\setcounter{equation}{0}

Another important difference between general relativity and quantum gauge theory of
gravity is the definition of an inertial reference system. It is known that an inertial reference
system is a special kind of coordinate system in which the equations of motion can hold in
their usual form. In classical mechanics, an inertial reference system is at rest in
absolute space-time, or in a state of uniform motion with respect to absolute space-time.
In general relativity, this definition is essentially changed. In general relativity, a local
inertial reference frame is considered to be a kind of inertial reference frames.
As stated above, the physics nature of the equivalence principle is to define clock
and ruler, or say that  to define space-time metric. Under this definition, gravity
disappears in this local system, space-time becomes local flat, and the equations
of motion can hold in their usual form in this local space-time. From the point of
view of a local observer, this local reference frame is almost the same as the
inertial reference frame. But for an observer in absolute space-time, even if in
local space-time, the local inertial reference frame is not an inertial reference
frame, for in the local inertial reference frame, there exist both gravity and
inertial force, the observer in it falls freely, and the clock and ruler used by
him are changed continuously. So, for an observer of absolute space-time,
the equations of motion can not hold in their usual form in this local system.
Of course, if we redefine clock and ruler at each point of space-time, or say
that redefine metric of space-time, we can make gravity and inertial force
exactly cancel each other at each point in the local space-time, and the
equations of motion returns to their usual forms in the local space, which
is just the treatment of general relativity.
\\
What determines the inertial reference system? What is the origin of inertial
force? In fact, quantum gauge theory of gravity can not answer such kinds of
questions. In quantum gauge theory of gravity, absolute space-time is introduced
transcendentally, or say that, it is the logical starting point of the whole theory.
According to definition, an inertial reference system is at rest in
absolute space-time, or in a state of uniform motion with respect to absolute
space-time. This definition means that an inertial reference system is also
introduced into the theory transcendentally. They are also logical starting point
of physics. We can consider this question from another point of view. Suppose
that there is a free mass point, which has no physical interactions with
exterior environments. According to energy-momentum conservation law,
the energy-momentum of the mass point must be a constant, or say
that, its velocity is always a constant. So, it must be at rest or in a state of
uniform motion. Before we make this conclusion, we must select a reference
system first. But, if we select a non-inertial reference system, we can not
obtain such conclusion. We will find that the mass point moves arbitrary, and
its energy-momentum changes arbitrary. For the observer in the non-inertial
reference system, the magnitude of this change can not be predicted, and
the cause of this change is also unknown. In this case, it is difficult to construct
a scientific kinematical theory. It is known that, in order to construct a scientific
theory, a key requirement is that there should be no effect if there is no
cause. In other words, for a free mass point, if it has no interactions with
exterior environments, its energy-momentum should be a constant. According
to this requirements, the only reference system that we can select is an inertial
reference system. After an inertial reference system is defined, how to
understand the non-inertial reference system? In fact, if we do not allow to
change our clock and ruler freely, we will find that there does not exist
any non-inertial reference system in nature. In other words, the physical
nature of a non-inertial reference system is that the observer in the non-inertial
reference system uses variable clock and ruler. For an observer in absolute
space-time, the observer in a non-inertial reference system is in a state of
variable motion. According to special relativity, the clock and ruler used
by the observer in a non-inertial reference system are changed when velocity
is changed. When an observer uses these variable clock and ruler in the
measurement, he will find that his reference system is a non-inertial
reference system. Suppose that there is a mass point which is at rest in a
non-inertial reference system. An observer in absolute space-time will
find that the mass point moves in variable velocity, and its energy-momentum
varies continuously. According to classical mechanics, this mass point must
feel a force from exterior environments. According to Newton's third law
of motion, there must be a reacting force, which is the inertial force observed
by the non-inertial observer. From the point of view of reference frame
transformation, the inertial force is  a physical effect of a residual term
of reference frame transformation. Therefore, inertial force is not a kind of
real physical interactions. Essentially speaking, the physical origin of inertial
force is the using of variable clock and variable ruler in the observation. Or say
that, inertial force is a residual physical effect of variable clock and variable
ruler. It is not a kind of physical interactions.
\\
What determines an inertial reference system? In fact, such kind of questions
is beyond the research field of physics, for an inertial reference frame is the
transcendental foundation of both classical physics and modern physics.
Up to now, there does not exist a physical theory of a non-inertial reference frame.
Discussions on inertial itself can only be performed philosophically.
If we discuss it in physics, we will discuss an inertial frame by using results
of an inertial frame, which is a logic ring. In fact, an inertial frame is the most
natural frame in physics. When we perform physical study, we need firstly
remove all possible interference from exterior environments, and isolate the
object of our study from exterior environments. The object should have no
physical interactions and no energy-momentum exchange with exterior
environments. This isolated system must be at rest in absolute space-time,
or in a state of uniform motion with respect to absolute space-time. The frame
fixed in this system is the inertial frame. In fact, any frame which fixed in
a free mass point is an inertial frame.
\\

\section{Logic System}
\setcounter{equation}{0}

In general relativity, because of the equivalence principle, a local inertial frame
is considered to be a kind of inertial frame, which make the concepts of
space-time, inertial and gravity are entangled with each other. In quantum
gauge theory of gravity, the concept of gravity is independent from
the concepts of space-time and inertial. Gravity is treated as a kind of fundamental
interactions. This is the important perceptional difference between two
theories. The goal and motivation of quantum gauge theory of gravity
are to study quantum effects of gravitational interactions. In order to study
quantum gravity, it is important to separate gravity from structures of
space-time, for we can not put quantum graviton into the structure of
space-time. There are two reasons. One reason is that we can not select
a local inertial frame, in which the quantum graviton disappears. Or say that,
we can not select a local inertial frame, in which all effects of quantum
gravity are shielded. Another reason is that the probabilistic quantum
wave function of quantum gravitational field can not be put into the metric
of space-time.
\\
We notice that the origin of the great difference between two theories
is from the equivalence principle.  For the equivalence principle, different
theorists have different interpretations on its physics role. The idea of the
equivalence principle comes from the consideration of the physics in a
free fall elevator. In a free fall elevator, the inertial force and gravity exactly
cancel each other, which is the physical reason that finally give birth to
the equivalence principle. Because of the equivalence principle, the local
inertial reference frame is regarded as a kind of inertial reference frame, which
can be explained in another way that all effects of gravity are completely
shielded in the local inertial reference frame. In fact, such kind of explanation
is not correct, and is not logically necessary. Strictly speaking, not all effects of
classical gravity can be completely shielded in the local inertial reference
frame. Only effects of classical gravito-electric field can be shielded in the local
inertial reference frame, effects of classical gravito-magnetic field can not be
shielded. A classical example is the spin-spin interactions\cite{17,18,19,20,21}.
A satellite which is in
an orbit around the earth can be regarded as a free falling frame.
Owing to the spin-dependent force, the gyroscope is neither relatively static, nor
uniform rectilinear moving, but oscillating with increasing amplitude. The existence
of this oscillation means that the gyroscope feels gravitational force in the free
falling frame, which violates the weak equivalence principle. However,
what the effects of classical gravity can be observed in a local inertial frame does
not violate the rationality of general relativity. As we have stated before,
the physical nature of the equivalence principle is to define clock and ruler,
or the space-time metric. In fact, for the logic system of general
relativity, the principle of general covariance is relatively more important and
more fundamental. In the construction of general relativity, the principle of
general covariance plays a key role, which brings gravity into mathematical
formulation of general  relativity. From the point of view of quantum gauge
theory of gravity, the principle of general covariance is the geometrical version
of gauge principle. The logic system of general relativity is as below
\begin{enumerate}
\item the clock and ruler, or the metric of space-time, are selected by
defining a local inertial reference frame,
\item gravity is introduced into the mathematical formulation of general relativity
through the principle of general covariance, and base on it, the physics in curved
space-time is set up,
\item the space-time metric, or the gravitational field in space-time,  is calculated
by solving Einstein's field equation.
\end{enumerate}
The logic system of quantum gauge theory of gravity is as below
\begin{enumerate}
\item transcendentally introduce an absolute space-time, and define the corresponding
clock and ruler, which are not affected by gravitational field in space-time,
\item the gravitational field in space-time is determined by solving the field equation of
gravitational gauge field,
\item the interactions between gravity and ordinary matter field are determined by
gauge principle.
\end{enumerate}

  .

\end{document}